\documentclass[fleqn,twoside]{article}
\usepackage{espcrc2}
\usepackage[dvips]{graphics}

\usepackage{graphicx}
\usepackage[figuresright]{rotating}

\newcommand{\AmS}{{\protect\the\textfont2
  A\kern-.1667em\lower.5ex\hbox{M}\kern-.125emS}}

\title{Phase Fluctuation of Fermion Determinant in Lattice QCD at Finite Density}

\author{Yuji Sasai\address{Oshima National College of Maritime Technology, 
 	Yamaguchi 742-2193, Japan       
	}%
        \thanks{Presenter}, 
        Atsushi Nakamura\address{RIISE, Hiroshima University, Higashi-Hiroshima 739-8421, Japan}
 	and 
	Tetsuya Takaishi\address{Hiroshima University of Economics, Hiroshima 731-0192, Japan}}
       
\begin{document}

\begin{abstract}
Once the quark chemical potential $\mu$ is introduced in finite density QCD, 
the fermion determinant appeared in the path integral measure becomes complex. 
In order to investigate the phase effect of $SU(3)$ lattice QCD (2-flavors), 
we calculated the fluctuation of the phase of $\det\Delta(\mu)$ on a $8^3\times4$ lattice at $\mu = 0.1$ and $0.2$. 
Then we calculated the chiral condensate and Polyakov line in the no phase and reweighted cases.
There is little difference between these two cases at $\mu = 0.1$ and $0.2$. 
We consider a possible reason for this result below $\mu \le 0.27$ in terms of $Z_3$ symmetry.
\vspace{1pc}
\end{abstract}

\maketitle

\section{Introduction}
Finite density QCD~\cite{MNNT03} has attracted considerable attention in high energy physics, nuclear physics and astrophysics. Many theorists now believe that QCD has a very rich structure when we study it in temperature and density parameter space, and some experimentalists want to reveal it.

Fodor and Katz\cite{Fodor02} studied QCD at $T\ne0$ and at $\mu\ne0$ by
the multiparameter reweighting method.
If two measures at $\mu=0$ and $\mu\ne0$ overlap well,
then the method works well.
However, for large $\mu$ the method may not work 
due to the phase effect.
Better overlap may be obtained 
by a simulation with a phase-quenching measure\cite{FKT02}. 
Here the difference between two measures is a complex phase $e^{i\theta}$.
If the phase fluctuates very much, then the method again fails,
which is expected to happen for large $\mu$.

It is therefore important to investigate the behavior of the
complex phase in $(T,\mu)$ parameter space.
In this study, we calculate the complex phase for various $\beta$ and $\mu$
and  study the phase fluctuation.
Furthermore, we use the complex phase to obtain 
results at finite $\mu$.

\section{Numerical Investigation}
The lattice partition function for KS fermions is given by 
\begin{equation}
  Z =\int\! DU (\det\Delta(\mu))^{N_f/4} e^{-\beta S_g}.
\end{equation}
We use the phase-quenching measure $\displaystyle \sim DU$ $|\det\Delta(\mu)|^{N_f/4} e^{-\beta S_g}$
in our Monte Carlo simulations.
We perform calculations on a $8^3\times4$ lattice 
at $m = 0.05$ for $N_f = 2$ using the R-algorithm.
The trajectory length is set at $0.5$ with a step size of $\Delta\,t=0.01$.
The first $1000$ trajectories were discarded for thermalization.

For $N_f=2$, the expectation value of an operator $O$ is given by
\begin{eqnarray}
 &\langle O \rangle& \!\!\!\!\!
  = \frac{1}{Z} \int\! DU (\det\Delta(\mu))^{1/2} O e^{-\beta S_g}
                                                            \nonumber \\
 \!\!\!\!\!\!&=&\!\!\!\!\!\!
    \frac{\int\! DU |\det\Delta|^{1/2} e^{i\theta /2} O e^{-\beta S_g}}
    {\int\! DU |\det\Delta|^{1/2} e^{i\theta /2} e^{-\beta S_g}}    \nonumber \\
 \!\!\!\!\!\!&=&\!\!\!\!\!\!
    \frac{\int\! DU |\det\Delta|^{1/2} e^{i\theta /2} O e^{-\beta S_g}}
    {\int\! DU |\det\Delta|^{1/2} e^{-\beta S_g}}                   \nonumber \\
 \!\!\!\!\!\!& &\!\!\!\!\!\!\!\!
    /\frac{\int\! DU |\det\Delta|^{1/2} e^{i\theta /2} e^{-\beta S_g}}
    {\int\! DU |\det\Delta|^{1/2} e^{-\beta S_g}} 
    \!= \!\frac{\langle O e^{i\theta /2}\rangle_0}
      {\langle e^{i\theta /2}\rangle_0},
                                                      \label{eq:reweight}
\end{eqnarray}
where $\langle\ldots\rangle_0$ is the expectation value with the phase-quenching measure. 

To obtain the phase $\theta$ of the fermion determinant, it is necessary to calculate the fermion determinant. 
We consider two methods for calculating $\det\Delta(\mu)$ : (I) calculation of eigenvalues, (II) LU factorization. 
We found that LU factorization is much faster than method (I).
Thus we have decided to use method (II) for the phase calculations.
\begin{figure}[t]
\includegraphics[scale=0.42, clip]{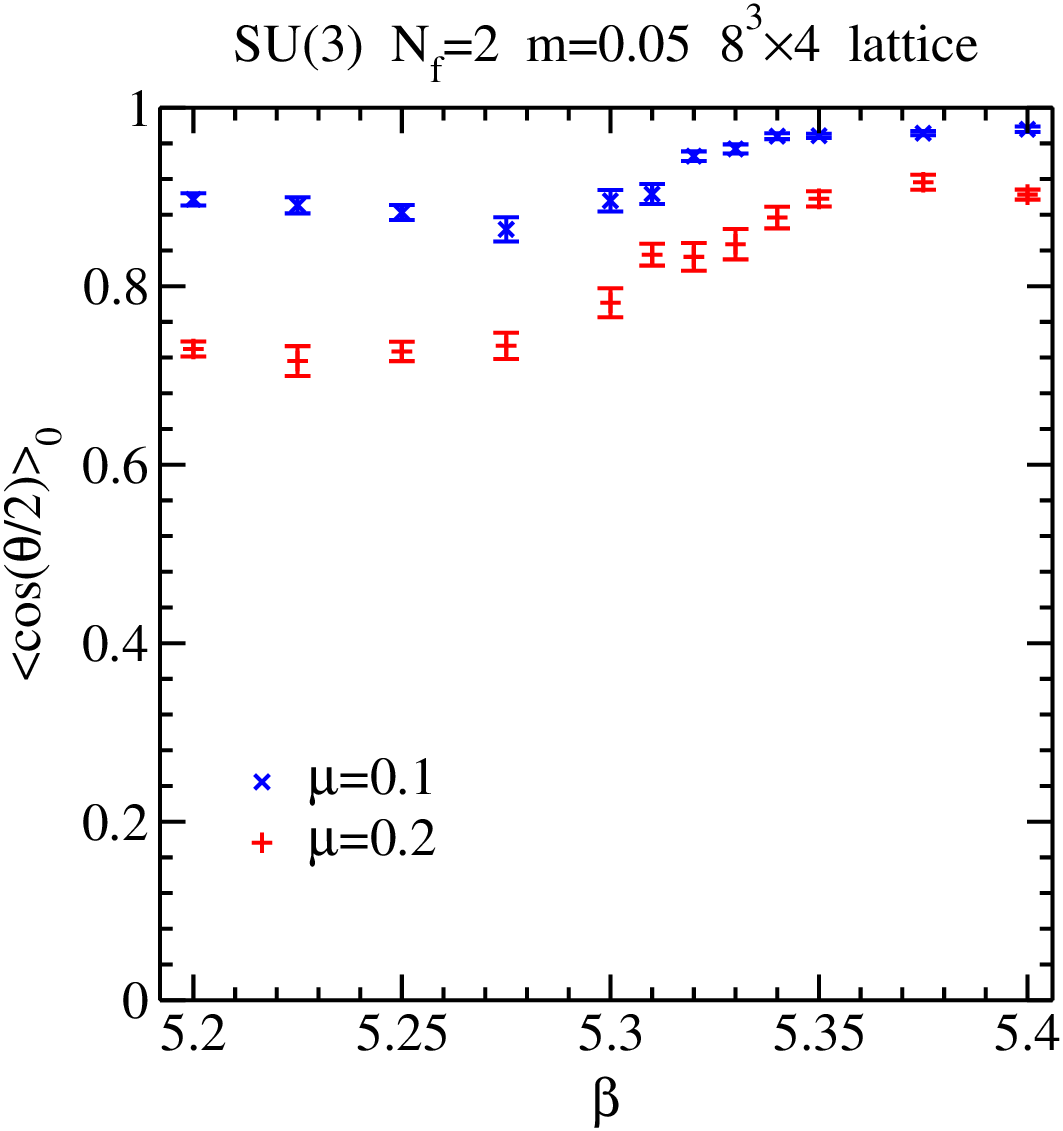}
\caption{$\langle \cos(\theta/2)\rangle_0$ for $\mu = 0.1$ and $0.2$, as functions of $\beta$}
\label{cos}
\end{figure}
Figure~\ref{cos} shows the fluctuations of $\langle \cos(\theta/2)\rangle_0$.  
The fluctuations increase with increasing $\mu$. 
The fluctuations decrease for $\beta > \beta_c$, where 
critical coupling $\beta_c$ is around $5.3$. 

Using the phase $\theta$ of $\det\Delta(\mu)$, we estimate the reweighted values of the chiral condensate $\langle {\rm Re}\,\bar\psi\psi\rangle$ 
according to reweighting formula (\ref{eq:reweight}), i.e.,
\begin{eqnarray}
\langle {\rm Re}\,\bar\psi\psi\rangle =
   \frac{\langle {\rm Re}\,\bar\psi\psi \,\cos(\theta/2)\rangle_0}
        {\langle \cos(\theta/2)\rangle_0}.
\end{eqnarray}
\begin{figure}[t]
\includegraphics[scale=0.42, clip]{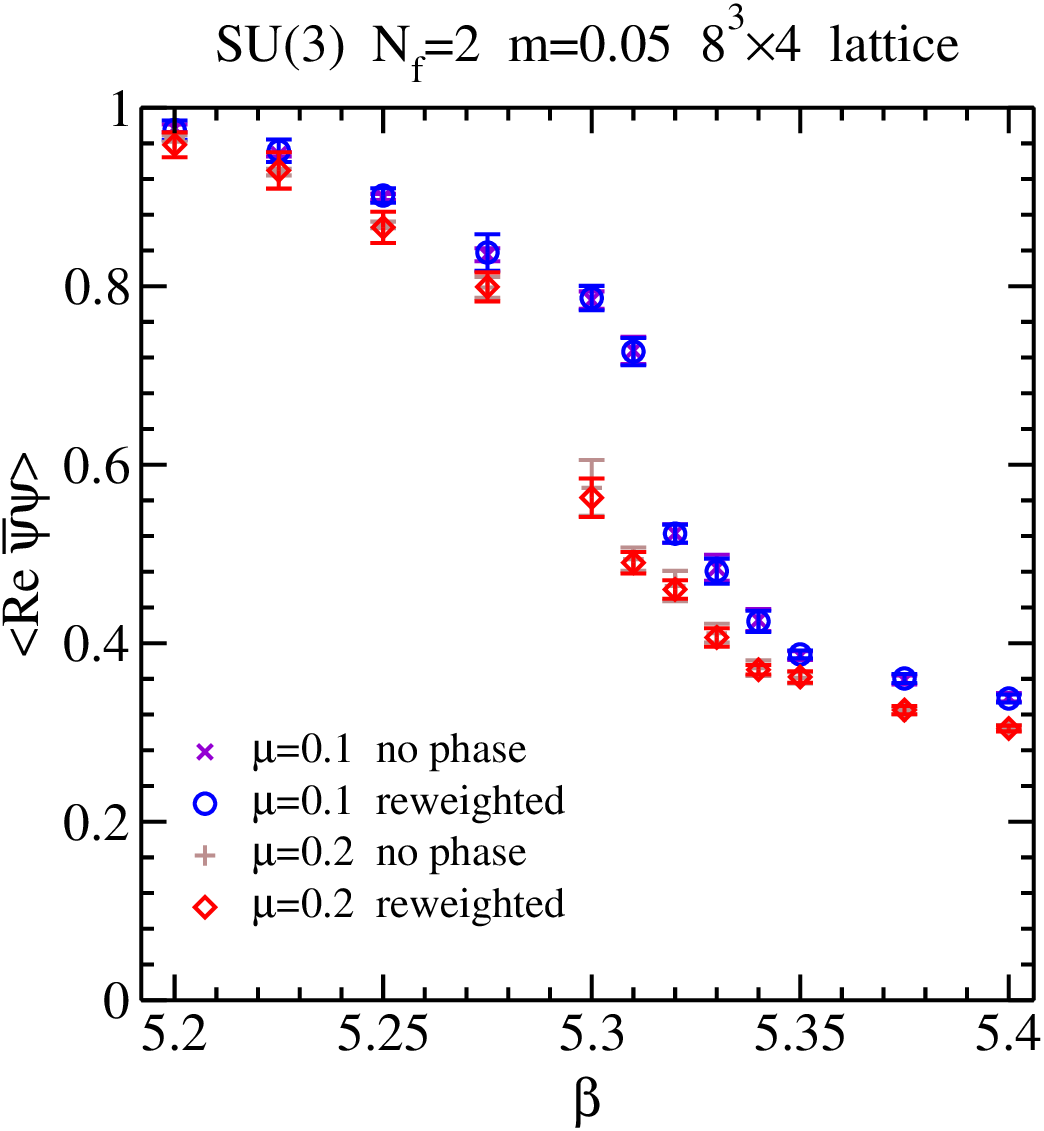}
\caption{No phase and reweighted chiral condensates $\langle {\rm Re}\,\bar\psi\psi\rangle$ for $\mu = 0.1$ and $0.2$, as functions of $\beta$}
\label{chiral}
\end{figure}
Figure~\ref{chiral} shows the graphs of the no phase case and the reweighted case. 
Values for the reweighted case are very similar to those for the no phase case. 
The latter agrees with the $\mu$ behavior reported by Kogut and Sinclair~\cite{Kogut02}. 
\begin{figure}[htb]
\vspace{2pt}
\includegraphics[scale=0.42, clip]{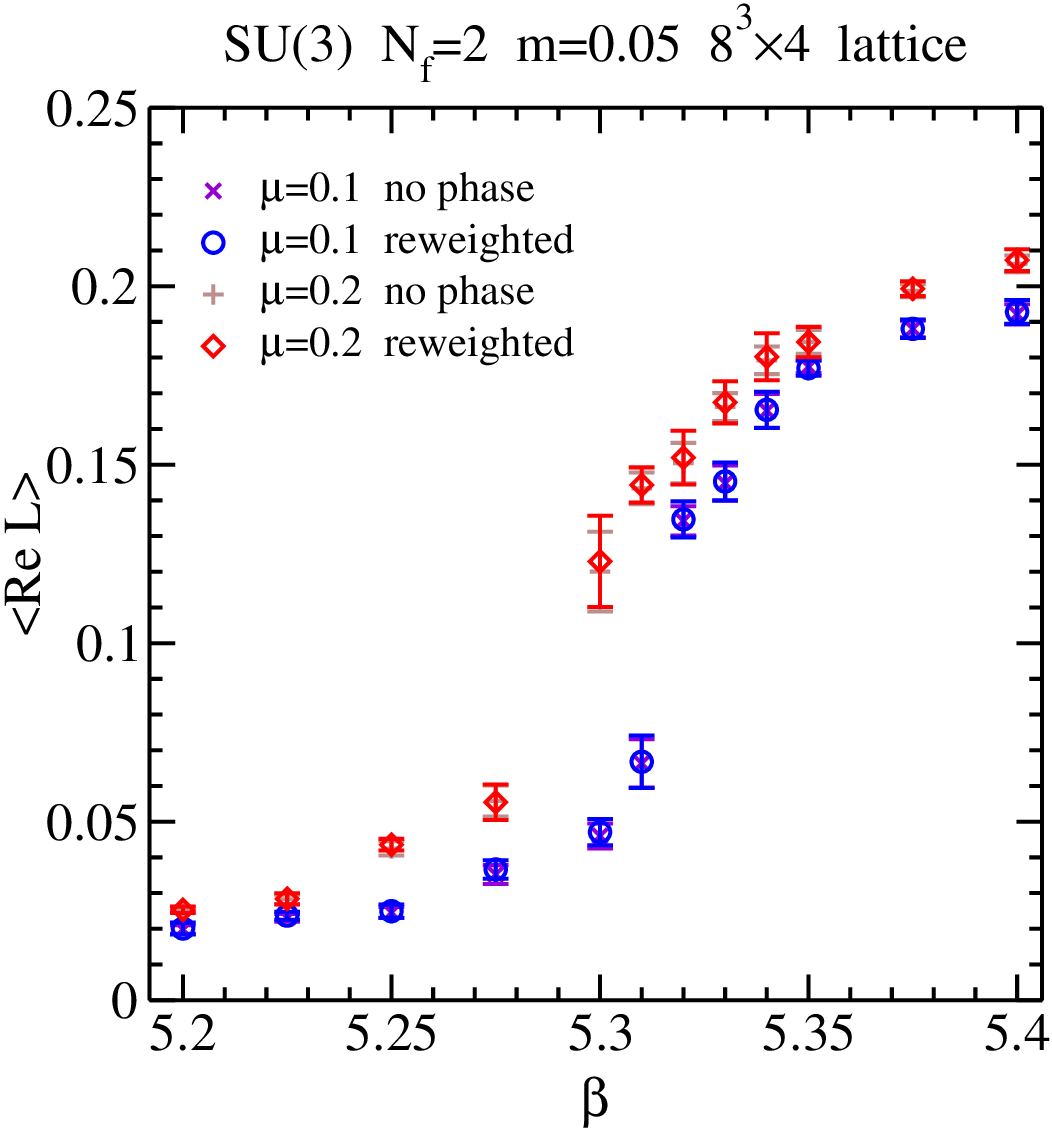}
\caption{No phase and reweighted Polyakov lines $\langle {\rm Re}\,L\rangle$ for $\mu = 0.1$ and $0.2$, as functions of $\beta$}
\label{polyakov}
\end{figure}

Figure~\ref{polyakov} shows Polyakov lines for the no phase case and the reweighted case. The two are also almost the same. 
In our approach, the conjugate gradient method does not converge at $\mu > 0.27$ and $\beta = 5.20$. 
We are performing simulations below $\mu_c$. 
If we can perform simulations above $\mu_c$, 
we might see the difference between the two cases.

\section{$Z_3$ symmetry}
Why does the phase effect not appear? 
Aarts et al.~\cite{Aarts02} proposed a partial summation in configuration $\{U\}$ connected by $Z_3$ symmetry. Let us write a set of link variables $\{ZU\}$ after the $Z_3$ transformation, i.e., we multiply $z = exp(2i\pi/3) \in Z_3 = \{1, e^{2i\pi/3}, e^{4i\pi/3}\}$ to all $\{U_4(x)\}$ on a time slice. Since $z$ is an element of $SU(3)$, $D(ZU) = DU$. The gauge action is invariant, $S_G(ZU) = S_G(U)$. Therefore,
\begin{eqnarray}
 Z \!\!\!\!&=&\!\!\!\!
       \int\! D(ZU) (\det\Delta(ZU))^{1/2} e^{-\beta S_g(ZU)} \nonumber \\
   \!\!\!\!&=&\!\!\!\!
       \int\! DU (\det\Delta(ZU))^{1/2} e^{-\beta S_g(U)}.
\end{eqnarray}
The same is true for $\{Z^2U\}$. Then we obtain
\begin{eqnarray}
 Z \!\!\!\!&=&\!\!\!\!
       \frac{1}{3}\int\! DU \left(\det\Delta(U)\right)^{1/2}  \nonumber \\
   \!\!\!\!& &\!\!\!\! 
       \biggl(1+ \Bigl(\frac{\det\Delta(ZU)}{\det\Delta(U)}\Bigr)^{1/2}
         + \Bigl(\frac{\det\Delta(Z^2U)}{\det\Delta(U)}\Bigr)^{1/2}\biggr)
                                                              \nonumber \\
   \!\!\!\!& &\!\!\!\! e^{-\beta S_g(U)}.
\end{eqnarray}
For any configuration $\{U\}$, $\{ZU\}$ and $\{Z^2U\}$ will appear 
in a whole set of ensembles.
Therefore, it is worth calculating the following partial summation in configuration $\{U\}$ connected by $Z_3$ symmetry:
\begin{eqnarray}
 E \!\!\!\!&\equiv&\!\!\!\! 
     1+ \Bigl(\frac{\det\Delta(ZU)}{\det\Delta(U)}\Bigr)^{1/2}
     + \Bigl(\frac{\det\Delta(Z^2U)}{\det\Delta(U)}\Bigr)^{1/2} \nonumber \\
   \!\!\!\!&=&\!\!\!\! |E|\,e^{i\theta}.
\end{eqnarray}
It is interesting to see whether the phase disappears in $E$. 
We are now trying to calculate this quantity.

\section{Summary}
We calculated the fluctuation of the phase of $\det\Delta(\mu)$ on a $8^3\times4$ lattice at $\mu = 0.1$ and $0.2$. In the region of small $\beta$, the fluctuations are large.
We calculated the chiral condensate $\langle {\rm Re}\,\bar\psi\psi\rangle$ and the Polyakov line $\langle {\rm Re}\,L\rangle$ in both no phase and reweighted cases. 
There is no difference between them. 
We are currently running at $\mu = 0.25$, varying $\beta$ through the crossover region.

We are trying to recognize these no phase effects below $\mu \le 0.27$ in terms of $Z_3$ symmetry. For more larger $\mu$, it is expected that this reweighting method will be effective. We must try to calculate the case of $\mu > 0.27$. The calculations of gluon energy density and fermion number density are also in progress now.


\begin{thebibliography}{9}
\bibitem{MNNT03} S. Muroya, A. Nakamura, C. Nonaka and T. Takaishi, 
                 hep-lat/030631 (2003).
\bibitem{Fodor02} Z. Fodor and S.D. Katz, JHEP 0203 (2002) 014; hep-lat/0106002.
\bibitem{FKT02} Ph. de Forcrand, S. Kim, T. Takaishi, Nuclear Physics B(Proc. Suppl.) 119 (2003) 538; 
                 hep-lat/0209126.
\bibitem{Kogut02} J.B. Kogut and D.K. Sinclair, Phys. Rev. D66, 034505 (2002); 
                 hep-lat/029054.
\bibitem{Aarts02} G. Aarts, O. Kaczmarek, F. Karsch and I.-O. Stamatescu, 
                 hep-lat/0110145.
\end{thebibliography}
\end{document}